\documentclass[graybox]{svmult}

\usepackage{type1cm}        
\usepackage{makeidx}         
\usepackage{graphicx}        
\usepackage{multicol}        
\usepackage[bottom]{footmisc}
\usepackage{algorithm}
\usepackage{algorithmic}
\usepackage{soul}
\usepackage{newtxtext}       %
\usepackage[varvw]{newtxmath}       

\makeindex             

\begin{document}

\title*{Functional regression with randomized signatures: An application to age-specific mortality rates}
\titlerunning{Functional regression with randomized signatures} 
\author{Zhong Jing Yap\orcidID{0009-0000-5599-8515},\\ Dharini Pathmanathan\orcidID{0000-0002-3279-4031} and\\ Sophie Dabo-Niang\orcidID{0000-0002-4000-6752}}
\institute{Zhong Jing Yap \at Institute of Mathematical Sciences, Faculty of Science, Universiti Malaya, 50603, Kuala Lumpur, Malaysia, \email{sim190030@siswa.um.edu.my}
\and Dharini Pathmanathan \at Institute of Mathematical Sciences, Faculty of Science, Universiti Malaya, 50603, Kuala Lumpur, Malaysia, \email{dharini@um.edu.my}
\and Sophie Dabo-Niang \at UMR8524–Laboratoire Paul Painlevé, Inria-MODAL, University of Lille, CNRS, Lille, 59000, France;\\
CNRS–Université de Montréal, CRM–CNRS, Montréal, Canada, \email{sophie.dabo@univ-lille.fr}}
%
%
\maketitle

\abstract*{We propose a novel extension of the Hyndman-Ullah (HU) model to forecast mortality rates by integrating randomized signatures, referred to as the HU model with randomized signatures (HUrs). Unlike truncated signatures, which grow exponentially with order, randomized signatures, based on the Johnson–Lindenstrauss lemma, are able to approximate higher-order interactions in a computationally feasible way.  Using mortality data from four countries, we evaluate the performance of the novel HUrs model compared to two alternative HU model versions. Our empirical results show that the proposed HUrs model performs well, particularly for Bulgarian and Japanese data.}

\abstract{We propose a novel extension of the Hyndman-Ullah (HU) model to forecast mortality rates by integrating randomized signatures, referred to as the HU model with randomized signatures (HUrs). Unlike truncated signatures, which grow exponentially with order, randomized signatures, based on the Johnson–Lindenstrauss lemma, are able to approximate higher-order interactions in a computationally feasible way.  Using mortality data from four countries, we evaluate the performance of the novel HUrs model compared to two alternative HU model versions. Our empirical results show that the proposed HUrs model performs well, particularly for Bulgarian and Japanese data.}

\section{Introduction}
Over thirty years since the introduction of the Lee-Carter (LC) model, the mortality forecasting literature has seen a steady rise in new methodologies that adopt its extrapolation approach. Some notable extensions include adjustments to the time-index \cite{LeeMiller2001}, addition of age-cohort interactions \cite{Booth2002}, \cite{RenshawHaberman2003} and \cite{Currie2013} employed the LC methodology within a generalised linear model framework, and \cite{LiLee2005} extended the LC model to forecast multiple populations. Recent literature also see the deployment of neural networks to forecast mortality as seen in \cite{RichmanWuthrich2021,SchnurchKorn2022,MarinoLevantesiNigri2022}. See \cite{BaselliniCamardaBooth2023} for a recent review on the LC model.

Another well-known extension is the Hyndman-Ullah (HU) model \cite{HU}, which uses the functional data paradigm \cite{RS} and applies nonparametric smoothing to view the mortality curves as functional time series, as well as functional principal component analysis (FPCA) to decompose the mortality curves. It distinguishes itself from the LC model by allowing multiple principal components to capture broader variation within the data, relaxes the homoskedasticity assumption of the LC model by allowing the error variance to vary with age, and employs more advanced time series methods in place of the random walk with drift model to forecast the time index. 
\cite{HU} reported superior point forecast accuracy over the LC model when forecasting French mortality. Variants of the HU model that have been proposed include a robust version \cite{HU}, and a weighted version (wHU) \cite{wHU}.

Recently, \cite{fmr} integrated signature methods from rough path theory into the HU model by replacing the FPCA decomposition with regression using truncated signatures, which is referred to as the HU with truncated signatures (HUts) model. This was motivated by \cite{ferm} who noted that signatures are naturally adapted to vector-valued functions, and require mild assumption on the regularity of functions to encode nonlinear geometric information. These properties suggest signature regression to be well suited for extending the HU model's forecasting framework. The HUts model leverages the universality of signatures by representing the mortality data into a series of signature coefficients that capture the underlying linear and nonlinear interactions inherent within the data. As a result, the HUts model demonstrated remarkable robustness in forecasting mortality rates, particularly in situations where data irregularities may arise due to disease outbreaks or periods of war. 

In this study, we investigate the utilization of an alternative signature representation based on linear random projections, specifically known as \textit{randomized signatures}. 

\section{Signatures and Randomized Signatures}
Signatures, rooted in the theory of rough paths, provide a systematic way to capture the geometric features of a time series or a functional trajectory through iterated integrals. Given a path  $X: [0,T] \to \mathbb{R}^d$, its signature can be thought of as an infinite series of terms.

\begin{definition}[Signature of a path \cite{ferm}]
    \label{DefSig}
  \textit{Let $X:[0,T]\rightarrow \mathbb{R}^d$ be a path of bounded variation, and let $I=(i_1,\ldots,i_k)\in\{1,\ldots,d\}^k$, where $k\in\mathbb{N}^*$, be a multi-index of length $k$. The signature coefficient of $X$ corresponding to the index $I$ is defined by}
  \begin{equation}
    \label{SigCo}
    S^I(X) = \idotsint\limits_{0 \leq u_1 < \dots < u_k \leq T} dX_{u_1}^{i_1} \dots dX_{u_k}^{i_k} 
  \end{equation}
\textit{where $S^I(X)$ is called the signature coefficient of order $k$. The signature of $X$ is then the sequence containing all the signature coefficients:}
 \begin{equation}
    \label{SigPath}
    S(X)=(1, S^{(1)}(X), \ldots, S^{(d)}(X), S^{(1,1)}(X), \ldots, S^{(i_1,\ldots,i_k)}(X), \ldots). \nonumber  
 \end{equation}
\end{definition}

The first term is always 1 (by convention). Subsequent terms are iterated integrals of the path increments, capturing increasingly higher-order interactions. 

\subsection{Truncated Signatures}
While the signature is infinite-dimensional in theory, practical usage requires truncation to a fixed level $m$, denoted as:
 \begin{equation}
    \label{SigPathTrunc}
    S^m(X)=(1, S^{(1)}(X), \ldots, S^{(d)}(X), S^{(1,1)}(X), \ldots, S^{(i_1,\ldots,i_m)}(X)).
\end{equation}
\noindent This is called the \textit{ truncated signature of order $m$}. As higher order iterated integrals often become progressively smaller or less informative, the bulk of the geometric information about the path is captured by lower order terms, making the truncated signature an efficient local approximation.

\subsection{Randomized Signatures}
The randomized signature is a dimensionality reduced version of the classical path signature. While the truncated signature encodes a path through a selected series of iterated integrals to capture its geometric features, one of its main drawbacks is the number of signature coefficients grows exponentially with the truncation order. In a high-dimensional settings or situations where rich higher-order information is critical, this exponential increase becomes computationally infeasible. Conversely, opting for a lower truncation order may overlook important nonlinearity and higher order interactions in the data. Randomized signatures alleviate this computational burden by applying a random projection to the high-dimensional signature space, thereby obtaining a compact, yet highly expressive, feature representation.


At the core of randomized signatures lies the Johnson–Lindenstrauss (JL) lemma, a powerful result in high-dimensional geometry. The JL lemma ensures that high-dimensional data can be projected into a lower-dimensional space with minimal distortion. More precisely, we have the following result:

\begin{lemma}[Johnson–Lindenstrauss Lemma \cite{paul}]
For every $0<\epsilon<1$ and every set $Q$ consisting of $N$ points in $\mathbb{R}^d$, there is a linear map $f: \mathbb{R}^d \to \mathbb{R}^k$ with $k> \frac{4\log N}{3\epsilon^2-2\epsilon^3}$ such that
\[
(1-\epsilon)\|u-v\|^2 \le \|f(u)-f(v)\|^2 \le (1+\epsilon)\|u-v\|^2
\]
for all $u,v\in Q$.
\end{lemma}

This property is crucial when reducing the dimensionality of the signature. Even after random projection, the essential geometric structure of the path is maintained. The definition of the randomized signature is as follows:

\begin{definition}[Randomized Signature \cite{com}]
\textit{Given $k \in \mathbb{N}$ and random matrices $A_1, \ldots, A_d \in \mathbb{R}^{k \times k}$, random shifts $b_1, \ldots, b_d \in \mathbb{R}^k$, random starting point $z \in \mathbb{R}^k$, and any fixed activation function $\sigma$, the randomized signature of $X$ at $t \in [0, T]$ is the solution of the differential equation:}
\begin{equation}
    dZ_t = \sum_{i=1}^d \sigma(A_i Z_t + b_i) \, dX_t^i, \quad Z_0 = z \in \mathbb{R}^k.
\end{equation}
\end{definition}

The random projection defined by these matrices is justified by the JL lemma, ensuring that the lower-dimensional embedding $Z_t$ retains the approximation properties of the full signature.

The randomized signature maintains the universal approximation property of the full signature \cite{cuchi1}. Recent theoretical results have demonstrated that, under appropriate conditions, one can approximate the evolution of controlled differential equations arbitrarily well by regressing on features generated via randomized signatures. Error bounds derived from the JL lemma provide explicit guarantees: if the projection dimension $k$ is chosen sufficiently large, then the distortion induced by the random projection is kept within a pre-specified tolerance \cite{paul}. This balance between reduced dimensionality and controlled approximation error is key to applying signature methods in high-dimensional settings.

The computational benefits of randomized signatures are twofold. First, by reducing the number of coefficients that need to be computed, they mitigate the exponential growth inherent in truncated signatures. Second, this reduction results in lower model complexity, which in turn accelerates training.These properties make randomized signatures especially attractive in machine learning tasks such as time series classification, system identification, and forecasting.

See \cite{com,cuchi1,cuchi2,bia} for a rigorous account on randomized signature.

\section{Methodology}

\subsection{The Hyndman-Ullah with Randomized Signatures Model}

The proposed Hyndman-Ullah with randomized signatures (HUrs) model extends the original HU framework by replacing the computation of truncated signatures in the HUts model with randomized signatures. This modification leverages the JL lemma to efficiently approximate higher-order interactions while mitigating the exponential growth of signature coefficients.

\subsubsection{Randomized Signature Algorithm}
Algorithm~\ref{alg:randomized_signature} details the procedure to generate the randomized signature of a path $X \in \mathbb{R}^d$.

\begin{algorithm}
\caption{Generate randomized signature \cite{com}}
\label{alg:randomized_signature}
\begin{algorithmic}[1]
\REQUIRE $X \in \mathbb{R}^d$ sampled at $0 = t_0 < t_1 < \dots < t_N = T$, randomized signature dimension $k$, activation function $\sigma$.
\STATE Initialize $Z_{t_0} \in \mathbb{R}^k$, and for each $i \in \{1, \dots, d\}$, set $A_i \in \mathbb{R}^{k \times k}$ and $b_i \in \mathbb{R}^k$ with i.i.d. standard normal entries.
\FOR{$n = 1$ to $N$}
    \STATE Compute:
    \[
    Z_{t_n} = Z_{t_{n-1}} + \sum_{i=1}^d \sigma(A_i Z_{t_{n-1}} + b_i) \left(X^i_{t_n} - X^i_{t_{n-1}}\right)
    \]
\ENDFOR
\end{algorithmic}
\end{algorithm}

\subsubsection{HUrs Model Procedure}
The steps to forecast age-specific mortality rates $y_t(x)$ with the HUrs model are as follows:

\begin{enumerate}
    \item \textbf{Smoothing:} Smooth the observed mortality rates using constrained weighted penalized regression splines to estimate the underlying mortality curves $f_t(x)$:
    \[
    y_t(x) = f_t(x) + \sigma_t(x) \epsilon_t,
    \]
    where $\sigma_t(x)$ is the age-dependent error variance and $\epsilon_t$ is a standard normal error term.
    
    \item \textbf{Mean adjustment:} Estimate the mean function $\mu(x)$ and compute the mean-adjusted rates:
    \[
    f^*_t(x) = \hat{f}_t(x) - \hat{\mu}(x).
    \]
    
    \item \textbf{Signature embedding:} For each age $x_i$, embed the time series $\{f^*_t(x_i)\}_{t=1}^n$ using basepoint augmentation, time augmentation, and the lead-lag transformation \cite{fmr}. Then calculate its randomized signature as per Algorithm~\ref{alg:randomized_signature}.
    
    \item \textbf{Principal component regression:} Apply principal component regression with the computed randomized signatures as predictors and $f^*_t(x)$ as the response. This decomposes the mortality curve into
    \[
    f_t(x) = \mu(x) + \sum_{k=1}^K \beta_{t,k} Z_k(x) + e_t(x),
    \]
    where $\beta_{t,k}$ are the time-dependent coefficients, $Z_k(x)$ is the $k$th principal component obtained via interpolation, and $e_t(x)$ is the model error.
    
    \item \textbf{Forecasting:} Fit univariate ARIMA models to the coefficients $\beta_{t,k}$ to forecast $\beta_{t+h,k}$ for a forecasting horizon $h$. Reconstruct the forecasted mortality rates $y_{t+h}(x)$ using the forecasted coefficients.
\end{enumerate}

The Hyndman-Ullah with truncated signatures (HUts) model follows the same procedure, except that in step 3 truncated signatures are computed in lieu of randomized signatures.

\section{Finite sample properties}

In this section, we assess the finite sample performance of the HUrs model by applying it to empirical mortality data. The evaluation is based on forecasting accuracy across multiple countries and forecasting horizons, enabling a comprehensive comparison with alternative models. First, we describe the mortality datasets used in our analysis, which span different countries, time periods, and age ranges. Following the data description, we present a detailed analysis of point forecasts, measured by the mean squared error (MSE), to demonstrate the performance of the proposed model.
\subsection{Data description}
\begin{figure}[!t]
\centering
\includegraphics[width=0.8\textwidth]{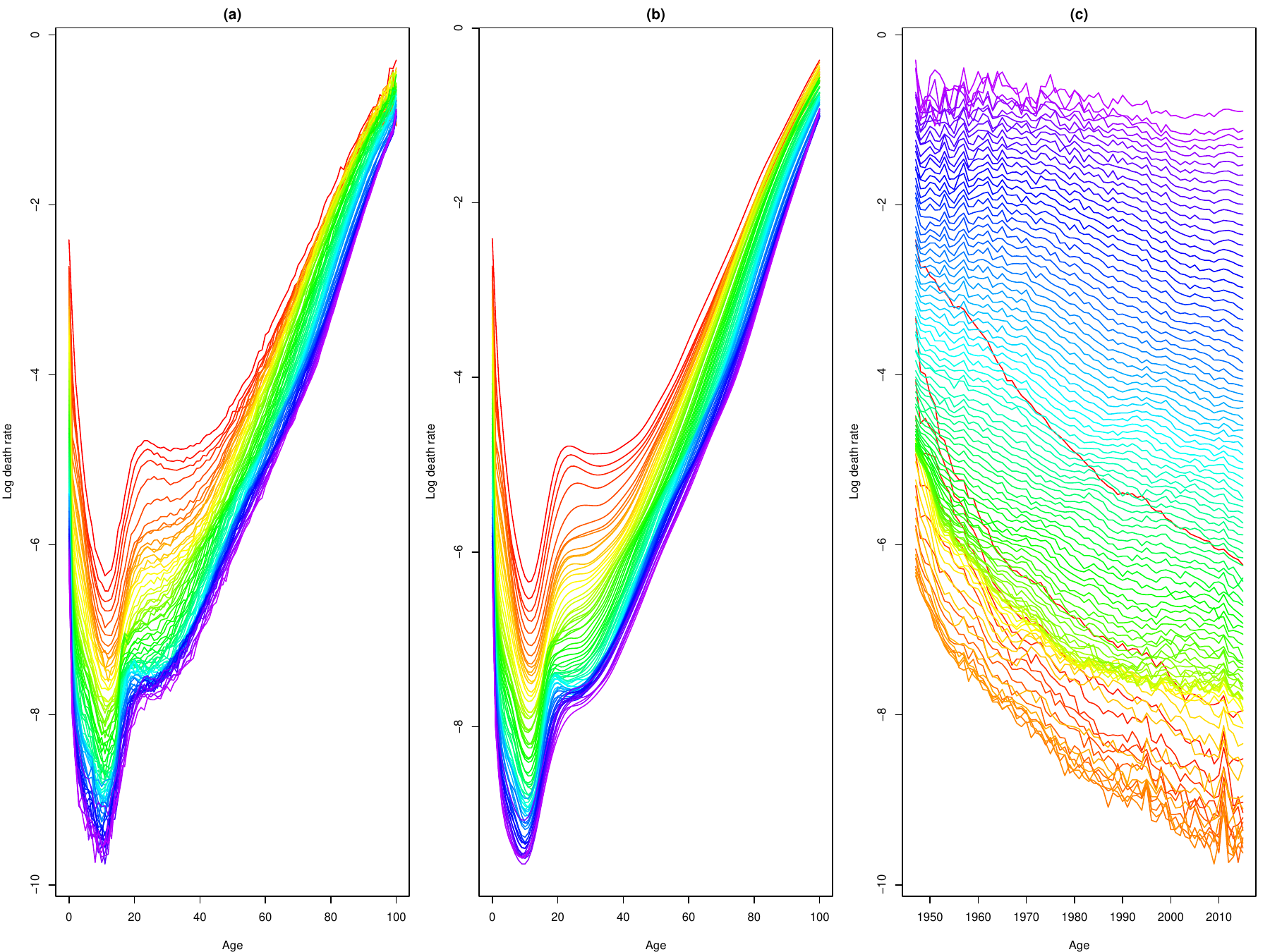}
\caption{Japan (1947 to 2015); (a) Observed log mortality rates; (b) Smoothed log mortality rates; (c) Time series for ages.}
\label{fig:JapanPLOT}
\end{figure}

Mortality data from four countries are obtained from the Human Mortality Database \cite{HMD}. Age groups were created, ranging from zero to a carefully determined maximum age to avoid zero or missing values in elderly data. For ages beyond this established maximum, they were combined into an upper age group. These are tabulated in Table \ref{tab:data_descn}. Figure \ref{fig:JapanPLOT} illustrates the log mortality rates of Japan.

\begin{table}[!t]
\caption{Mortality data by country}
\label{tab:data_descn}       
\begin{tabular}{p{3cm}p{3cm}p{2.2cm}}
\hline\noalign{\smallskip}
Country & Commencing Year & Age Range \\
\noalign{\smallskip}\svhline\noalign{\smallskip}
Belgium       & 1920--2015 & 0--100+ \\
Bulgaria      & 1947--2015 & 0--100+ \\
Japan         & 1947--2015 & 0--100+ \\
United States & 1933--2015 & 0--100+ \\
\noalign{\smallskip}\hline\noalign{\smallskip}
\end{tabular}
\end{table}

\subsection{Point forecast}
The mortality rates of four countries were forecasted using the HUrs model with $k=100$ and a linear activation function $\sigma=\frac{1}{d\sqrt{k}}$. Treatment of the path embeddings will result with a path dimension, $d$ of 3 as in \cite{fmr}. Consequently, it is intuitive to compare the forecast accuracy of the proposed method with the HUts model and the weighted HU (wHU) model, which were the best performing models in \cite{fmr}.

The models were fitted using an expanding window approach, incorporating data from each country’s beginning years as outlined in Table \ref{tab:data_descn}. The initial forecasting period covers the last two decades, ending at 2015. We fit the data with both models and produce forecasts up to ten years ahead. We evaluate the accuracy of these forecasts. Then, we extend the training period by a year and recalculate the forecasts. This process continues until 2014.

To assess the accuracy of the HUrs model’s point forecast, mean squared errors (MSE) are computed and compared to those of the HU model variants. MSE quantifies the average squared difference between predicted and observed values, indicating the overall accuracy and precision of predictions. Lower MSE values indicate better forecasting performance. The MSE is calculated using the following equation:
\begin{equation}\label{eq:mse_mape_h}
        MSE(h) = \frac{1}{pq}\sum^q_{t=1}\sum^p_{i=1}(y_t(x_i)-\hat{y}_{t|t-h}(x_i))^{2},
\end{equation}
where $h$ represents the forecasting horizon, $p$ denotes the number of ages, $q$ signifies the number of years considered, $y_t(x_i)$ denotes observed log mortality, and $\hat{y}_{t|t-h}(x_i)$ represents the predicted value. The results are tabulated in Tables \ref{tab:MSEHURs_h1}, \ref{tab:MSEHURs_h5} and \ref{tab:MSEHURs_h10}.

\begin{table}[!t]
\caption{MSE: HUrs, HUts, and wHU across different countries for $h=1$}
\label{tab:MSEHURs_h1}       
\begin{tabular}{p{2.5cm}p{2.2cm}p{2.2cm}p{2.2cm}}
\hline\noalign{\smallskip}
\textbf{Country} & \textbf{HUrs} & \textbf{HUts} & \textbf{wHU} \\
\noalign{\smallskip}\svhline\noalign{\smallskip}
Belgium       & 0.02371 & 0.02183 & \textbf{0.02047} \\
Bulgaria      & \textbf{0.01930} & 0.01942 & 0.01968 \\
Japan         & 0.00680 & \textbf{0.00674} & 0.00687 \\
United States & 0.00162 & 0.00133 & \textbf{0.00129} \\
\noalign{\smallskip}\hline\noalign{\smallskip}
\end{tabular}
\end{table}

\begin{table}[!t]
\caption{MSE: HUrs, HUts, and wHU across different countries for $h=5$}
\label{tab:MSEHURs_h5}       
\begin{tabular}{p{2.5cm}p{2.2cm}p{2.2cm}p{2.2cm}}
\hline\noalign{\smallskip}
\textbf{Country} & \textbf{HUrs} & \textbf{HUts} & \textbf{wHU} \\
\noalign{\smallskip}\svhline\noalign{\smallskip}
Belgium       & 0.04172 & 0.02627 & \textbf{0.02620} \\
Bulgaria      & \textbf{0.04442} & 0.04595 & 0.05478 \\
Japan         & \textbf{0.01000} & 0.01045 & 0.01020 \\
United States & 0.01023 & \textbf{0.00746} & 0.00896 \\
\noalign{\smallskip}\hline\noalign{\smallskip}
\end{tabular}
\end{table}

\begin{table}[!t]
\caption{MSE: HUrs, HUts, and wHU across different countries for $h=10$}
\label{tab:MSEHURs_h10}       
\begin{tabular}{p{2.5cm}p{2.2cm}p{2.2cm}p{2.2cm}}
\hline\noalign{\smallskip}
\textbf{Country} & \textbf{HUrs} & \textbf{HUts} & \textbf{wHU} \\
\noalign{\smallskip}\svhline\noalign{\smallskip}
Belgium       & 0.09131 & \textbf{0.0364} & 0.0404 \\
Bulgaria      & \textbf{0.08949} & 0.0976 & 0.1280 \\
Japan         & \textbf{0.03289} & 0.0348 & 0.0362 \\
United States & 0.03197 & \textbf{0.0226} & 0.0305 \\
\noalign{\smallskip}\hline\noalign{\smallskip}
\end{tabular}
\end{table}

The result demonstrates that different models excel for different countries and forecasting horizons. Specifically, the HUrs model consistently outperforms others for Bulgaria across all three forecasting horizons. Additionally, in Japan, the HUrs model shows strong performance for both $h=5,10$. These findings highlight the importance of selecting appropriate models tailored to each country’s unique characteristics and the specific forecasting periods.
\section{Conclusion and discussions}

In this study, we present a new extension of the Hyndman-Ullah (HU) model for forecasting mortality rates by incorporating randomized signatures. We assess the finite sample properties of the proposed model in comparison with two alternative versions of the HU model. The HUrs model demonstrates robust performance across the selected countries, particularly in Bulgaria and Japan. Furthermore, the modular design of the HUrs methodology indicates significant potential for optimization. Prospective advancements might involve experimenting with diverse embeddings, fine-tuning the $k$ values, and evaluating various activation functions to enhance their adaptability and performance across different scenarios.

Furthermore, beyond its application in mortality forecasting, the randomized signature method shows significant potential for tasks involving high-dimensional data. In such applications, it can greatly reduce computational costs and processing time, making it a valuable tool for efficiently handling complex datasets.

\ethics{Competing Interests}{
The authors have no conflicts of interest to declare that are relevant to the content of this chapter.}

\eject

\end{document}